\documentclass[journal]{IEEEtran}

\usepackage[utf8]{inputenc} 
\usepackage[T1]{fontenc}
\usepackage{color}
\usepackage{float}
\usepackage{amsbsy}
\usepackage{amstext}
\usepackage{setspace}
\usepackage{orcidlink}
\usepackage{graphicx}
\usepackage{bm}
\usepackage{float}
\usepackage{amsmath, amssymb}
\usepackage{gensymb}
\usepackage{upgreek}
\usepackage{breqn}
\usepackage{blindtext}
\usepackage{packages/widetext}
\usepackage{packages/dblfloatfix}
\usepackage{subcaption}

\begin{document}

\title{Threshold Properties in Degenerate Distributed Feedback Lasing}

\author{Albert~Herrero-Parareda~\orcidlink{https://orcid.org/0000-0002-8501-5775},
        Nathaniel~Furman~\orcidlink{https://orcid.org/0000-0001-7896-2929},
        and~Filippo~Capolino~\orcidlink{https://orcid.org/0000-0003-0758-6182}
\thanks{A.~Herrero-Parareda, N.~Furman and F.~Capolino are affiliated with the Department of Electrical Engineering and Computer Science, University of California, Irvine,
CA, 92697 USA. e-mail: f.capolino@uci.edu.}
}

\maketitle


\begin{abstract}
We investigate threshold properties in degenerate distributed feedback (DDFB) lasing in a double-grating photonic structure that operates near a degenerate band edge (DBE) with the potential to support robust single-frequency lasing. The DBE is an exceptional point degeneracy (EPD) of fourth order involving four coalescing Bloch eigenmodes. We also provide an analytical demonstration that, for a four-mode system, a  quartic flat Bloch dispersion implies the coalescence of the four eigenvectors into a single eigendirection, without explicitly calculating them. Unlike previous studies based on coupled-mode theory, here the realistic double grating is studied using a full-wave method. A DDFB photonic cavity operating close to the DBE frequency is shown to display a large quality factor that scales with the fifth power of the cavity length. 
Upon the inclusion of gain, the DDFB cavity displays a low lasing threshold with the exceptional scaling as the inverse of the fifth power of the double-grating length, i.e.,  $\alpha_{D,th} \propto 1/N^5$, where $N$ is the number of waveguide unit cells making the mirrorless cavity. These results suggest the potential for compact, single-frequency lasing using a mirrorless cavity with minimal device footprint. The DDFB laser is very attractive for various applications including communications, sensing, and spectroscopy.
\end{abstract}

\section{Introduction}
\label{ch:Intro}

The ability of lasers to generate coherent, monochromatic light makes them essential for high-speed data communications, sensing, and other cutting-edge technologies. Distributed feedback~(DFB) lasers are a type of laser diode widely used in these applications due to their stable single-frequency lasing mode operation and narrow linewidth \cite{Okai-DFB-JAP94,carroll1998distributed,  ghafouri_distributed_2003_ch1, morthier2013handbookDFB}. The DFB laser relies on a distributed feedback mechanism along a single-path waveguide, where the two propagating optical waves are coupled via a longitudinal grating (i.e., a corrugated structure along the direction of propagation of the guided waves) in the material. While the DFB mechanism provides a stronger frequency selectivity than conventional Fabry-Perot cavities, it tends to provide two lasing frequencies which are sensitive to perturbations in the temperature or in the boundary conditions of the structure. Usually, a defect is included in the cavity to provide increased stability \cite{loh_phase_1995, ghafouri_distributed_2003_ch3}. The DFB laser also requires introducing antireflective coatings on the sides of the DFB structure to suppress the Fabry-Perot modes in the cavity and avoid multi-mode lasing \cite{ghafouri_distributed_2003_ch2}. Furthermore, putting a mirror on one side of the DFB structure and an antireflective coating on the other side may create an unbalanced situation that would cause another frequency shift. Affordable DFB lasers with high performance are therefore challenging to produce due to the costly and complex fabrication processes of high-end laser diodes required for single-mode operation \cite{li_fiber_2009}. To enhance the performance of DFB lasers, this work proposes the concept of a degenerate DFB~(DDFB) laser. Based on the double-grating photonic structure introduced in \cite{mealy_degenerate_2023}, the DDFB lasing mechanism has the potential to lead to single-frequency lasing mode operation through a degeneracy between four different Bloch propagating and evanescent eigenmodes (whereas a DFB laser generally uses {\em two} propagating modes in the longitudinal grating). 

\begin{figure}
    \centering    
    \includegraphics[width=0.48\textwidth]{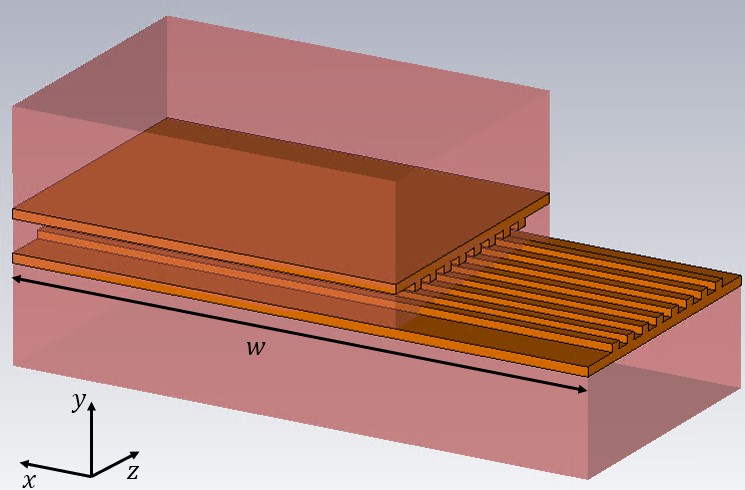}
    \caption{Schematic realization of the double grating making the mirrorless DDFB cavity with length of $N$ unit cells along $z$ and width $w$  (the term “mirrorless” indicates that no additional mirrors are required). The top part is truncated in the drawing to better show the three-dimensional structure. Also, $N$ in this figure is much smaller than $N$ considered in the rest of the paper.   At the DBE, the four Bloch modes polarized along $x$ and traveling along $z$ form a degenerate standing wave. The DBE resonance of the cavity occurs at a frequency very close to the DBE one.} 
    \label{fig:InfiniteDDFBStructure}
\end{figure}

The degenerate band edge (DBE) is a special exceptional point degeneracy (EPD) of  order four that occurs in a passive and lossless waveguide \cite{nada_theory_2017, veysi_degenerate_2018, abdelshafy_exceptional_2019, mealy2020generalConds,zheng2022synthesis}. EPDs are points in a system's parameter space at which the eigenvalues and eigenvectors collapse, or coalesce, on each other \cite{kato_perturbation_1966, heiss_exceptional_2004}. In the literature, EPDs in photonic systems have been studied mainly in waveguide with the presence of balanced gain and loss, as in parity-time (PT)-symmetric systems \cite{ruter_observation_2010, ramezani_unidirectional_2010}

\begin{figure}
    \centering
\includegraphics[width=0.48\textwidth]{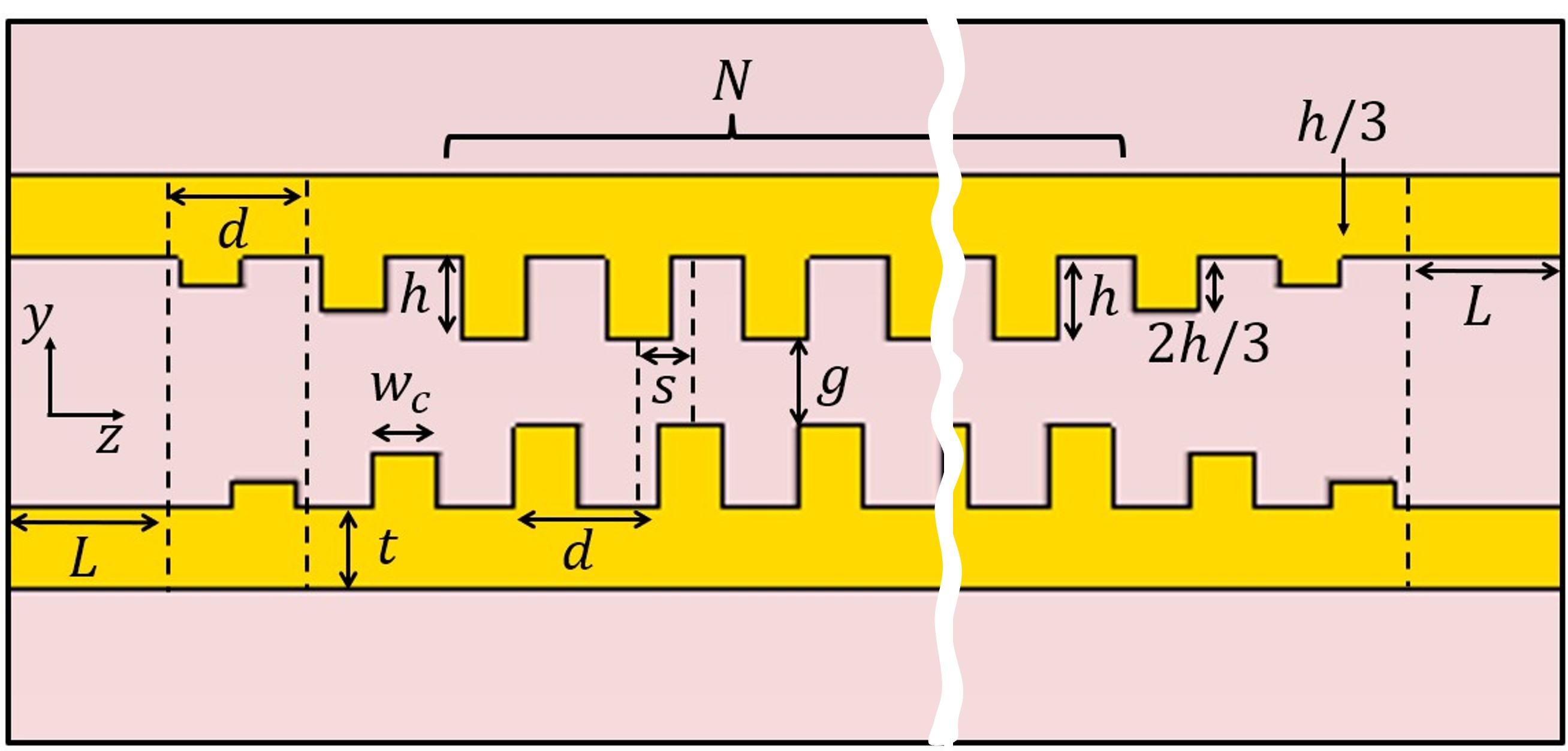}
    \caption{Double grating supporting guided longitudinal modes in the $z$ direction, polarized along $x$. The two oppositely-facing gratings are shifted by a distance $s$ along $z$ to break mirror symmetry, enabling the occurrence of the DBE that has a very flat band in the dispersion diagram provided by Eq.~(\ref{eq:DBEDispDiagr}). The tapering is performed by adding two cells at either end of the structure whose corrugation height decreases by $h/3$ consecutively. Termination regions of length $L$ are added beyond the tapering sections, and their interface is denoted by vertical dashed black lines. The double grating supporting only an RBE used for comparison purposes has mirror symmetry, i.e., $s=0$.} 
    \label{fig:DDFBStructure}
\end{figure}

The proposed active DDFB cavity is based on a waveguide supporting the DBE and by including distributed gain to bring instability in the cavity. Such a cavity displays a remarkably low lasing threshold that is here demonstrated to scale as the inverse fifth power of the waveguide length, and that lasing near the DBE frequency is preferred over lasing near the closest regular band edge in the same cavity. We additionally show that a DFB waveguide with the same structure and similar dimensions, supporting only a standing wave Bloch-mode degeneracy at a comparable frequency, exhibits a quality factor scaling as the cube of waveguide length and a lasing threshold scaling inversely with the cube of the length.

 The resulting DDFB lasing cavity does not require the introduction of defects within, and it provides a comparatively lower lasing threshold than the DFB cavity. Consequently, a DDFB laser shows promise in low-power optical telecommunication systems. In summary, the DDFB laser may offer several advantages over traditional DFB lasers due to the synchronized oscillation of the four degenerate waveguide modes. These potential advantages include enhanced light-matter interactions, narrower linewidth, reduced lasing threshold, and increased resilience to external perturbations. An analogous concept has been introduced at radio frequency in \cite{oshmarin2019IET-newOsc,Abdleshafy2020TAP-DistrDBEoscc,radfar2021IEEETCSlowPhaseNoise} and experimentally demonstrated in \cite{Oshmarin2021arXiv-experDBEoscillator}. The concept of DBE oscillator has been proposed also in vacuum electronics in \cite{othman_theory_2016,othman2016low,Chipengo2017Backward, Zuboraj2017Propagation,Abdelshafy2018Electron,MacLachlan2022Efficient, MacLachlan23TEDEffectsElectro,Maclachlan2024sub,saavedra2025double}.

 In addition to the full-wave investigation of DDFB lasing, we provide a mathematical demonstration of the connection between the quartic flat Bloch dispersion and the coalescence of four Bloch eigenvectors at a fourth-order EPD. Using the characteristic equation of a four-mode transfer matrix, we show that the corresponding eigenvalue has geometric multiplicity one and therefore is associated with a single fourth-order Jordan block. This demonstration, presented in the Appendix, does not require the explicit calculation of the eigenvectors and provides a rigorous mathematical foundation for identifying the DBE through its characteristic dispersion relation.

This paper is organized as follows: Section~\ref{ch:Double grating} defines the double-grating design, following the approach outlined in \cite{mealy_degenerate_2023}. In Section~\ref{ch:DBEinDDFB}, we introduce the design considerations and parameters for the waveguide to support a DBE in its modal dispersion relation and investigate its performance without gain. Section~\ref{ch:Lasing} provides the inverse fifth power scaling of the lasing threshold with cavity length when gain is introduced in the double grating, and how the lasing threshold depends on the reflection coefficient at the two cavity terminations. The performance of a future DDFB laser compared to that of a comparable DFB laser is discussed throughout the paper. The results are summarized in Section~\ref{ch:Conclusion}.

\section{Degenerate Distributed Feedback Waveguide}
\label{ch:DDFB-Waveguide}

\subsection{Degenerate Band Edge and Fourth-Order EPD}
\label{ch:DBE-EPD4}
A DBE in the Bloch dispersion relation of a periodic waveguide is a fourth-order EPD, i.e., it is characterized by the coalescence of four Bloch eigenvalues and their corresponding eigenvectors. Unlike a regular band edge (RBE), which involves a second-order EPD, the DBE exhibits a quartic dispersion relation near the degeneracy frequency, $(\omega_D-\omega) \propto  (k-k_D)^4$, resulting in a particularly flat dispersion diagram and vanishing group velocity \cite{figotin_gigantic_2005,figotin_frozen_2006, figotin2007slow-wave,sukhorukov2007slowLightDBE, jung2008photChrysDBE, sukhorukov2008slowDBE, yarga_degenerate_2008,  nada_theory_2017}.
Although EPDs are commonly associated with non-Hermitian systems involving gain and loss, they can also occur in passive, lossless periodic waveguides as demonstrated in \cite{abdelshafy_exceptional_2019,nada_theory_2017,nada_exceptional_2020,nada_frozen_2021,parareda_frozen_2022,nada2023designModCouplRes,furman_frozen_2023,kessem_low_2023,zamir2026soiDBE}. In these waveguides, the Bloch transfer matrix is generally non-Hermitian, even when the constituent materials are lossless. Under suitable geometrical and coupling conditions, this matrix becomes nondiagonalizable, allowing multiple Bloch eigenvalues and their corresponding eigenvectors to coalesce. Therefore, the occurrence of a DBE does not require material gain or loss \cite{abdelshafy_exceptional_2019,nada_theory_2017,nada_exceptional_2020,nada_frozen_2021,parareda_frozen_2022,nada2023designModCouplRes,furman_frozen_2023,kessem_low_2023,zamir2026soiDBE}. Note that, although the authors of Refs. \cite{figotin_gigantic_2005,figotin_frozen_2006} did not use the term exceptional point, they did provide the clear concepts of Jordan block and coalescence of eigenvectors. It has been demonstrated that a passive and lossless waveguide does not need to be periodic to possess an EPD, indeed the DBE and strategies to obtain it in uniform waveguides were demonstrated in \cite{mealy2020generalConds}.  The double-grating DFB waveguide presented here is periodic and an accurate analytical modeling would require several evanescent Floquet waves interacting between the two waveguides following either a procedure as in \cite{gutman_degenerate_2011,gutman2012slow-Frozen} or purely numerical as done here.  In this work, all the properties of the double-grating DDFB waveguide, including the quality factor and the lasing threshold are obtained using full-wave electromagnetic simulations based on the finite-element method implemented in CST Studio Suite. Despite the complicated coupling between the two gratings, each waveguide is designed to support {\em only two Bloch modes}, polarized along $x$, propagating in opposite directions. Cell-to-cell propagation of each Bloch mode is represented as $\mathbf T_{\mathrm U} \boldsymbol{\Psi}_n =\zeta_n\boldsymbol{\Psi}_n$,  where $\mathbf T_{\mathrm U}(\omega)$ is the unit-cell frequency-dependent transfer matrix, $\zeta_n=e^{-j k_n d}$, where $k_n$ is the Bloch wavenumber of the  $n$th mode ($n=1,2,3,4$),  $d$ is the waveguide period. The state eigenvector $\boldsymbol{\Psi}_n$ of the $n$th Bloch mode represents the weights of the transverse electric-field profiles of the selected Bloch mode. The frequency dispersion of the Bloch wavenumbers is found by solving  $\det\!\left(
\mathbf T_{\mathrm U}(\omega)- \zeta \mathbf I
\right)=0$ for complex $\zeta$. At the fourth-order EPD discussed in this paper, $\zeta_1=\zeta_2=\zeta_3=\zeta_4=-1$, that correspond to $k_D=\pi/d$.
Looking at the eigenvectors, it has been shown that this fourth-order degeneracy is verified when the Bloch modal dispersion diagram is such that $(\omega_D-\omega) \propto  (k-k_D)^4$ in a neighborhood of the DBE angular frequency $\omega_D$ \cite{figotin_gigantic_2005,nada_theory_2017,abdelshafy_exceptional_2019}.
In this work, the mathematical connection between the quartic dispersion and the coalescence of the four Bloch eigenvectors is established in the Appendix using the characteristic equation of a four-mode transfer matrix. In particular, under the nondegeneracy condition specified therein, we demonstrate that the fourfold Bloch eigenvalue has geometric multiplicity one: 
\begin{equation}
\dim\ker[\mathbf T_{ \mathrm U}(\omega_D)+\mathbf I]=1.
\end{equation}
Since the eigenvalue $\zeta_D=-1$ has algebraic multiplicity four but geometric multiplicity one, the transfer matrix $\mathbf T_{\mathrm U}$ has a single fourth-order Jordan block. Thus, all four eigenvectors $\boldsymbol{\Psi}_n$  coalesce into one degenerate eigendirection, establishing the connection between the quartic DBE dispersion and a fourth-order EPD.

The DDFB regime analyzed in this paper is associated with the formation of a DBE  in the Bloch modal dispersion  \cite{figotin_gigantic_2005, figotin_frozen_2006, gutman2012slow-Frozen, burr_degenerate_2013, wood_degenerate_2015, othman_giant_2016, nada_theory_2017, nada_giant_2018}. This exceptionally flat dispersion implies the vanishing group velocity $v_g=d \omega/ dk=0$ at the DBE, together with its derivatives $dv_g /dk=0$ and $d^2 v_g /dk^2=0$, with $d^3 v_g/dk^3\ne0$. 
Therefore, at resonances near the DBE frequency, suitable finite-length cavities exhibit exceptionally large group delays and loaded quality factors, both scaling asymptotically as the fifth power of the cavity length, i.e., $\tau_g\propto N^5$ and $Q\propto N^5$, where $N$ is the number of unit cells \cite{figotin_slow_2011, wood_degenerate_2015, nada_theory_2017,othman_giant_2016, othman_experimental_2017, veysi_degenerate_2018}.  Working at a resonance near the DBE also implies enhanced field amplitude \cite{figotin_slow_2011}, and enhanced local density of states \cite{othman_giant_2016}.  These properties provide the physical foundation for the DDFB lasing mechanism investigated in this work.

Based on previous theoretical studies on DBE-based lasing \cite{othman_giant_2016, veysi_degenerate_2018}, once an active material is included \cite{siegman_lasers_1986}, the DDFB photonic structure exhibits an enhanced interaction between the gain medium and the degenerate mode, which generates oscillations. The fundamental concept of leveraging  RBE effects to enhance lasing efficiency was analyzed in \cite{dowling_photonic_1994}. 
A preliminary concept of DBE-based lasing was presented in \cite{veysi_degenerate_2018} using an idealized coupled-mode model and gain mechanism, without specifying a realistic photonic structure operating at optical frequencies. Here, instead, we investigate DBE-based lasing in a more realistic double-grating photonic structure, using full-wave electromagnetic simulations to account for its actual geometry and cavity terminations.

\begin{figure}[htbp]
\centering
\includegraphics[width=0.42\textwidth]{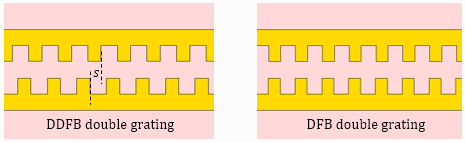} 
    \vspace{2mm} 
\includegraphics[width=0.5\textwidth]{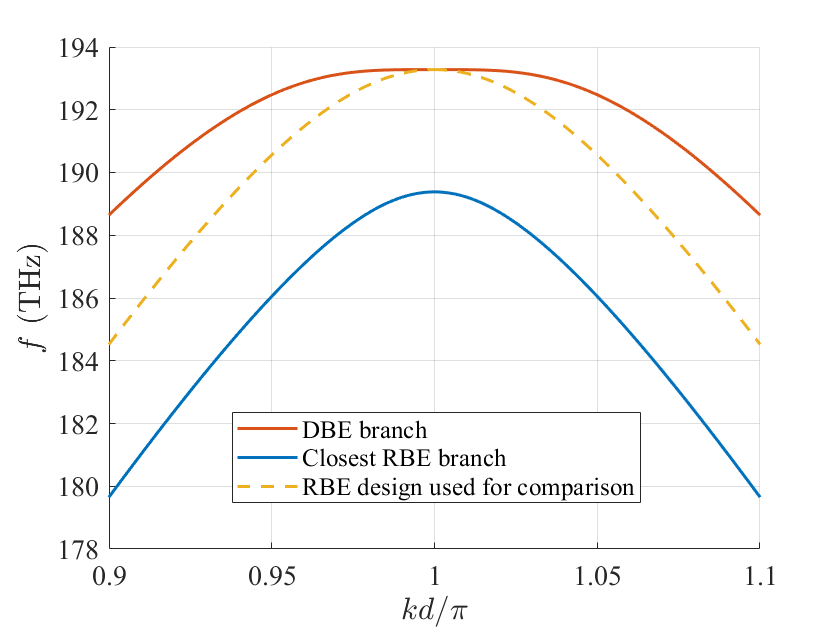}
\caption{Dispersion diagrams of the longitudinal modes for the DDFB and DFB double-grating designs. The DDFB double grating has two modal branches (orange and blue solid lines). The branch that includes the four-mode coalescence at the DBE frequency $f_D= 193.27$ THz is depicted as a continuous orange curve. The evanescent modes coalescing at the middle of the flat-band edge are not shown in the figure. The branch showing the closest RBE at $189.38$ THz is depicted as a continuous blue curve. The dashed yellow line represents the modal dispersion diagram of the DFB double-grating design with mirror symmetry (i.e., with $s=0$) that supports only an RBE at a frequency close to $f_D$, for comparison purposes. The flatness associated with the DBE (see Eq.~(\ref{eq:DBEDispDiagr})) is noticeably greater than those associated with the RBEs in the two designs.}
\label{fig:DispDiagr}
\end{figure}

\subsection{double-grating distributed feedback waveguide}
\label{ch:Double grating}

The DDFB photonic structure proposed here features a double-grating configuration where a standard grating is coupled with another grating. The second grating is obtained after performing a mirror operation with respect to the $x,z$ plane and a translation $s$ in the $z$ direction. The waves in each optical grating couple with the waves in the other grating, forming a coupled double-grating structure that supports four modes (when considering both directions). The translation $s$ breaks mirror symmetry, which would otherwise prevent the formation of a four-wave degeneracy (the DBE) in two coupled periodic waveguides. The example of the double grating, shown in Fig.~\ref{fig:InfiniteDDFBStructure}, is based on silicon-on-insulator~(SOI) technology, although it can be built with photonic materials (such as $\text{SiN}$, $\text{InP}$, $\text{LiNbO}_3$, $\text{GaAs}$, $\text{AlGaAs}$, etc). In periodic media, Bloch modes with small group velocities display a strong impedance mismatch with the external waveguide  connections \cite{povinelli_slow_2005, pottier_efficient_2007}. Hence, the coupling efficiency of light from an external medium into the slow-wave modes in the cavity is inversely proportional to the group velocity of the Bloch mode \cite{velha_compact_2007}. To slightly mitigate the abrupt discontinuities at the cavity terminations, and also to partially decrease radiation induced by an abrupt change in waveguide periodicity \cite{carin1993timeHarm,capolinoRS2009truncationWH}, we taper the last two unit cells on each side by reducing the corrugation height $h$ by one-third, as shown in Fig.~\ref{fig:DDFBStructure}. This tapering aims to reduce radiation losses caused by abrupt changes at the edges of the cavity and potentially improve coupling with external waveguides. The two unit cells involved in the tapering on each side are henceforth not included in the number of unit cells $N$ of the finite-length cavity.

Beyond the tapering, the cavity is connected to straight uniform waveguides with the same thickness $t$ as the rest of the slab. The junction point is denoted by the vertical dashed black lines in Fig.~\ref{fig:DDFBStructure}, which schematically depicts the double-grating configuration with tapers and terminations.

We show an example of a double-grating waveguide with a silicon core with a refractive index of $n_{Si}=3.45$, and a silicon dioxide cladding with a refractive index of $n_{SiO_2}=1.44$ \cite{barrios_demonstration_2007}. We assume the materials do not exhibit dispersive behavior in the small frequency range of interest \cite{dattner_analysis_2011}, that is, around the DBE frequency $f_D=193.27$ THz.

 For ease of numerical simulations using the FEM, we assume that the width $w$ of the coupled gratings is much larger than the wavelength, as shown in Fig.~\ref{fig:InfiniteDDFBStructure} and we assume also that the field is invariant in the  $x$ direction.  Although the physical waveguide width $w$ is assumed to be much larger than the wavelength, the field invariance along $x$ allows us to use a much narrower computational domain bounded by two perfect electric conductor~(PEC) walls orthogonal to $x$, thereby reducing the computational cost without significantly affecting the longitudinal modal properties.

Our objective is to determine the longitudinal modes that propagate in the $z$ direction and are polarized in the $x$ direction (using the coordinate system as shown in Figs.~\ref{fig:InfiniteDDFBStructure}~and~\ref{fig:DDFBStructure}). The Bloch dispersion relation of the longitudinal modes in the infinitely-long double-grating structure is calculated by solving the full-wave electromagnetic eigenvalue problem for a single periodic unit cell implemented in the FEM solver of CST Studio Suite. This is done by imposing a prescribed Bloch phase shift between the two longitudinal boundaries of a single unit cell. The eigenfrequencies are obtained for different real Bloch wavenumbers, allowing the dispersion branches near the Brillouin-zone boundary to be reconstructed. This procedure directly accounts for the spatial field variations within the corrugated unit cell and does not require constructing an approximate four-mode transfer matrix. The identification of the DBE is based on the characteristic quartic dispersion near the Brillouin-zone boundary.

The  transverse computational domain is terminated by four PEC walls, two orthogonal to the $x$ direction and two orthogonal to the $y$ direction. In the $y$ direction, the two PEC walls are sufficiently far from the double grating to avoid interference with the evanescent field in the cladding of the longitudinal modes. We verified this by varying the PEC wall distance from the waveguide core and confirming the results remain consistent. Additionally, we also performed checks using perfect magnetic conductor (PMC) boundaries instead, orthogonal to the $y$ direction, obtaining the same result.

\section{DBE resonance in a DDFB Cavity}
\label{ch:DBEinDDFB}

\subsection{DBE and RBE Dispersion Characteristics}
The lossless and gainless double-grating configuration supporting a DBE at $f_D$ (known as the DDFB waveguide or double grating) is found via a combination of manual and numerical parameter optimization where the second derivative of the dispersion, $d^2\omega/ dk^2$, of the Bloch modes at the edge of the Brillouin zone $k=\pi/d$ is minimized, i.e., $|d^2\omega/ dk^2|_{\pi/d} \to 0$. Minimizing the magnitude of this derivative ensures the Bloch degeneracy involves all four modes supported by the double grating shown in Fig.~\ref{fig:DDFBStructure}. Minimizing just the magnitude of the group velocity $|v_g|\to 0$ may result in the formation of an RBE instead. The continuous lines in Fig.~\ref{fig:DispDiagr} show the real-wavenumber branches of the modal dispersion diagram of the DDFB design with the following parameters (in nm): $t=81$, $s=119$, $d = 335$, $h = 102$, $g = 214$, and $w_c = 92$. The solid-orange curve denotes the modes that coalesce at the DBE frequency $f_D = 193.27$ THz with the typical  very flat band edge, whereas the solid-blue curve at lower frequencies denotes the modes that coalesce forming a regular band edge~(RBE) that occurs in the same double-grating structure. The RBE is a second-order degeneracy involving only the coalescence of two counter-propagating Bloch modes in the waveguide, which form a standing wave \cite{figotin_gigantic_2005}.  Typically, a DFB laser operates in the vicinity of an  RBE (generally referred to only as band edge) \cite{kogelnik_coupled_1972}, whereas the DDFB operates in close proximity to the DBE. Near the RBE frequency $f_R$, the dispersion diagram is approximated as \cite{figotin_gigantic_2005}

\begin{equation}
	(f_R-f) \approx \frac{h_R}{2\pi} (k-k_R)^2,
	\label{eq:RBEDispDiagr}
\end{equation}

whereas at frequencies near the DBE, the dispersion diagram is approximated as \cite{figotin_gigantic_2005,figotin_frozen_2006, gutman_frozen_2012, burr_degenerate_2013, nada_giant_2018}

\begin{equation}
	(f_D-f) \approx \frac{h_D}{2\pi} (k-k_D)^4,
	\label{eq:DBEDispDiagr}
\end{equation}

where $k_R$ and $k_D$ are the wavenumbers at the DBE and the RBE, respectively (in this case, it is $\pi/d$ for both cases). The parameters $h_R$ and $h_D$ indicate the flatness of the band edges and are sometimes called the ``flatness parameter''. In this case, the band edge separates a pass band from a stop band at higher frequencies; therefore, both $h_R$ and $h_D$ are positive. At the DBE, $d^n\omega/dk^n=0$ for $n=1,2,3$, and $d^4\omega/d_k^4=- 24 h_D$. Though not shown in Fig.~\ref{fig:DispDiagr} because only the propagation branches are shown, there are four modes coalescing at the DBE when considering also the evanescent modes. This can be easily seen by considering the four quartic roots

\begin{equation}
(k-k_D) \approx \left[2\pi (f_D-f)/h_D\right]^{1/4}
\end{equation} 
as clearly shown in \cite{gutman2012slow-Frozen, nada_theory_2017, veysi_degenerate_2018, abdelshafy_exceptional_2019, mealy_degenerate_2023}. Above the DBE frequency $f_D$, all four modes are evanescent for the case represented in Fig.~\ref{fig:DispDiagr}.


To compare the performance of DFB and DDFB cavities, we introduce another double-grating configuration that supports {\em only} an RBE (close to the DBE frequency of the DDFB design), and which is henceforth referred to as the DFB double grating or DFB cavity (shown on the top right of Fig.~\ref{fig:DispDiagr}). In this double-grating RBE design we impose mirror symmetry (by letting $s=0$), which prohibits the formation of the DBE, and use the same period $d = 335$ nm and comparable waveguide dimensions. The other parameters of the DFB design are (in nm): $t=80$, $h = 100$, $g = 350$, and $w_c = 80$. 
 The DFB double-grating configuration considered here is used as a reference structure to isolate the effects associated with the different orders of band-edge degeneracy. Both designs have the same period and comparable geometrical dimensions, but support a second-order RBE and a fourth-order DBE, respectively. This comparison is not intended to represent a performance benchmark against an optimized conventional single-grating DFB laser, such as a phase-shifted DFB design commonly employed in practical applications. Such lasers may achieve lower thresholds than those considered here through optimization of their coupling, gain distribution, and cavity design. Rather, our objective is to investigate the distinct resonance properties and asymptotic scaling laws associated with two comparable, not optimized structures supporting RBE and DBE resonances.

The real-branch modal dispersion diagram for the DFB double-grating structure is shown with the dashed-yellow line in Figure~\ref{fig:DispDiagr}. At the edge of the Brillouin Zone $k=\pi/d$, the dispersion of the DDFB design (solid lines) displays a much greater flatness than that relative to a regular DFB design (dashed lines), as seen by comparing the solid orange curve with the yellow dashed one.
The reason for the flatter modal dispersion around the DBE frequency compared to the RBE frequency in Fig.~(\ref{fig:DispDiagr}) is that the flatness at the RBE frequency is only due to the coalescence of two modes, whereas the flatness at the DBE frequency is associated with the two propagating waves and two evanescent waves composing the four degenerate modes \cite{figotin_electromagnetic_2013}. 

\subsection{Resonances and Quality-Factor Scaling}
Next, we consider a double-grating cavity made of $N$ unit cells. For the cavity to display the properties associated with the DBE, a cavity resonance must occur in proximity to the DBE frequency. We call the resonance closest to the DBE the ``DBE resonance'' with associated frequency $f_{r,D}$, as in \cite{yarga_degenerate_2008, burr_degenerate_2013, nada_theory_2017, nada_giant_2018, veysi_degenerate_2018}. 
The DBE resonance frequency of the cavity becomes closer and closer to the DBE frequency when increasing the cavity length, following the asymptotic law 

\begin{equation}
    f_{r,D} \sim f_{D} - \frac{h_D}{2 \pi} \left(\frac{\pi-\varphi}{N d}\right)^4,
    \label{eq:DBEresFreqAsymp}
\end{equation}

where $\sim$ denotes asymptotic equality for large $N$, as discussed in \cite{burr_degenerate_2013, othman_giant_2016, othman_experimental_2017, nada_theory_2017, nada_giant_2018, veysi_degenerate_2018}. The angle $\varphi$ represents a reflection phase of each propagating mode at the edges of the cavity, where also the evanescent waves are strongly excited \cite{burr_degenerate_2013}.  The smaller the flatness parameter $h_D$ is, the closer the DBE resonance frequency is to the DBE frequency. (Note that the asymptotic formula reported in \cite{othman_experimental_2017,nada_theory_2017,nada_giant_2018,veysi_degenerate_2018, abdelshafy_exceptional_2019} should have included also the term $\varphi$). Similarly, for the DFB cavity, the resonance closest to the RBE frequency is known as the ``RBE resonance'' with associated frequency 

\begin{equation}
f_{r,R} \sim f_{R}- \frac{h_{\mathrm{R}}}{2 \pi}\left(\frac{\pi}{Nd}\right)^2.
    \label{eq:RBEresFreqAsymp}
\end{equation}

 The different asymptotic scaling laws of the loaded quality factors for DBE and RBE cavities originate from their distinct Bloch dispersion relations and the associated finite-length resonance conditions. Near the DBE, the resonance condition in Eq.~\eqref{eq:DBEresFreqAsymp} implies that the resonance frequency approaches the DBE frequency as $f_{r,D}-f_{D}\propto N^{-4}$. However, the quality-factor scaling cannot be inferred from the resonance detuning alone, since it also depends on the coupling of the nearly degenerate Bloch modes and the boundary conditions at the cavity terminations. For suitable finite-length DBE cavities, the interaction of the four nearly degenerate Bloch modes leads to an asymptotic loaded quality-factor scaling $Q_D\propto N^5$, consistent with the theoretical framework developed in Refs. \cite{figotin_gigantic_2005,figotin2007slow-wave}. By contrast, the quadratic dispersion near an RBE leads to $Q_R\propto N^3$ under analogous conditions \cite{bendickson1996analytic}. Here, we investigate a more realistic double-grating photonic structure, whose electromagnetic properties and termination effects are determined through full-wave simulations rather than an approximate four-mode model. Our results verify the established quality-factor scaling laws for this specific geometry, including its tapering and terminations.

The resulting frozen mode in the DDFB cavity exhibits comparatively larger field amplitude enhancement and group delay compared to the RBE-associated slow wave resonance \cite{figotin_gigantic_2005, figotin2007slow-wave, figotin_slow_2011}. As a result, waveguides operating near the DBE often demonstrate enhanced performance compared to those operating near an RBE \cite{figotin_slow_2011, othman_giant_2016, nada_theory_2017, nada_giant_2018}. 
 
The existence of the DBE at optical frequencies has been experimentally demonstrated in silicon photonic waveguides, as reported in Refs.\cite{wood_degenerate_2015,reano_exp_2016}. These works confirmed the DBE by reconstructing the dispersion diagram and by observing the characteristic quality factor scaling $Q_D \sim N^5$ at the DBE resonance frequencies of cavities made by $N$ unit cells. Additional experimental evidence of the DBE has been reported at microwave frequencies  \cite{othman_experimental_2017, abdelshafy_exceptional_2019,mealy2020generalConds,zheng2022synthesis}.

The optical properties of the finite-length DFB and DDFB waveguides are obtained using the FEM solver in CST Studio Suite. 
In the following simulations, excitation is applied from the left side of the cavity using the even mode of the waveguides at a distance $L$ from the corrugations, as shown in Figure~\ref{fig:DDFBStructure}. The amplitude of the excited mode at the input port is scaled such that the total power carried by that mode equals $1$ W.

\begin{figure}
    \centering
    \includegraphics[width=0.48\textwidth]{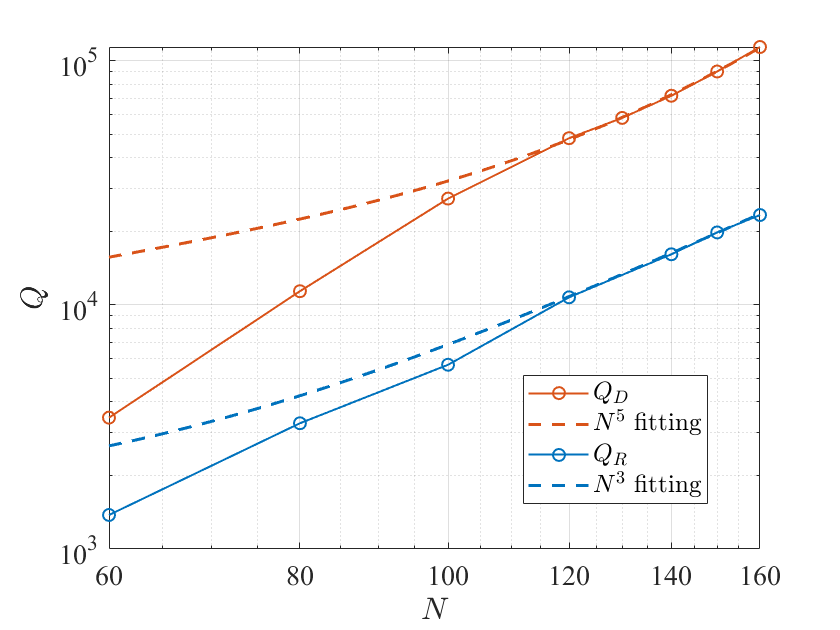}
    \caption{Comparison between the loaded quality factors of the DDFB (orange dots) and the DFB (blue dots) cavities (evaluated at their respective resonances) varying grating length. The dashed lines of the same color are the fitting functions that show the asymptotic $N^5$ scaling of the DBE quality factor $Q_D$ with DDFB waveguide length (orange) and the $N^3$ scaling of the RBE quality factor $Q_R$ with DFB waveguide length (blue). Both devices have the same period.}
    \label{fig:QvsN}
\end{figure}

The quality factors of the lossless and gainless DFB and DDFB cavities are shown in Figure~\ref{fig:QvsN} for different numbers of unit cells $N$. Both quality factors are computed for a loaded double-grating structure with tapering of the corrugations of two additional unit cells on either side and uniform waveguides with thickness $t$, extended continuously on both sides (without mirrors) as illustrated in Figure~\ref{fig:DDFBStructure}. The junction point is denoted by the vertical dashed black lines.

The quality factors are calculated as $Q= \pi f_r \tau_g(f_r)$, where $f_r$ is the resonance frequency for either the DDFB ($f_{r,D})$ or the DFB ($f_{r,R})$ cavity, and $\tau_g(f_r)$ is the group delay at that frequency calculated as \cite{othman_giant_2016, nada_theory_2017}

\begin{equation}
    \tau_g(f) = \frac{1}{2\pi}\frac{\partial\angle S_{21}}{\partial f},
    \label{eq:DefGroupDelay}
\end{equation}

where $\angle$ denotes the phase of the transmission $S_{21}$ from input to output. The transmission is calculated using two ports, one on each side of the resonator along $z$. Each electromagnetic port spans both gratings in the transverse cross section and extends into the cladding to account for the evanescent fields surrounding the silicon cores. Unless explicitly stated, in the following simulations, $L=0$ between the port location and the beginning of the coupled gratings.

In Figure~\ref{fig:QvsN}, the orange dots represent the quality factors of the DDFB cavity $Q_D$, while those of the DFB cavity $Q_R$ are shown as blue dots. The dashed lines of the same color represent their respective fitting trends; $Q_D \approx aN^5 + bN$ for the DBE, where $a=6.93\times 10^{-7}$ and $b=252.18$, and $Q_R \approx cN^3 + d$ for the DFB, where $c=0.0054$ and $d=1.478\times 10^3$. The first term of the $Q_D$ fitting function dominates when $N>138$, whereas that of the $Q_R$ fitting function dominates for $N>64$, demonstrating that these fittings describe the asymptotic scaling of the quality factor with waveguide length. The observed $N^5$ and $N^3$ asymptotic scaling laws are consistent with the literature for both degeneracies  \cite{figotin_slow_2011, wood_degenerate_2015, othman_giant_2016, nada_theory_2017, nada_giant_2018, othman_experimental_2017,abdelshafy_exceptional_2019,   mealy2020generalConds,zheng2022synthesis, zamir2026soiDBE}. Besides the two completely different scaling laws summarized in Table~\ref{tab:DDFB_DFB}, we also observe that the quality factor of the DDFB cavity is larger than that of the DFB cavity for any $N$, owing to the exceptional four-mode degeneracy. This result demonstrates that the DBE-associated frozen mode regime in the DDFB design experiences greater impedance mismatch at the cavity terminations compared to the RBE-associated resonance of the DFB design. Consequently, the DDFB laser is anticipated to achieve lasing with a lower lasing threshold without mirrors and to be more resilient to perturbations of its terminations than the DFB laser. This strong mismatch occurs because at the DBE resonance both the coalescing propagating and evanescent modes together are necessary to satisfy the boundary conditions at the two edges of the cavity. Since they have almost coincident polarization states because of the four-wave degeneracy  \cite{figotin_slow_2011, gutman_degenerate_2011}, both types of waves have to be large to reconstruct the field at the terminations. This phenomenon results in having propagating modes with large magnitude causing the field enhancement at the interior of the cavity, away from the terminations. This DBE-related phenomenon is substantially different from what happens in conventional uniform Fabry-Perot cavities or in a DFB cavity where there are no evanescent waves needed to satisfy the boundary conditions at the terminations.

\begin{figure}
    \centering
    \includegraphics[width=0.48\textwidth]{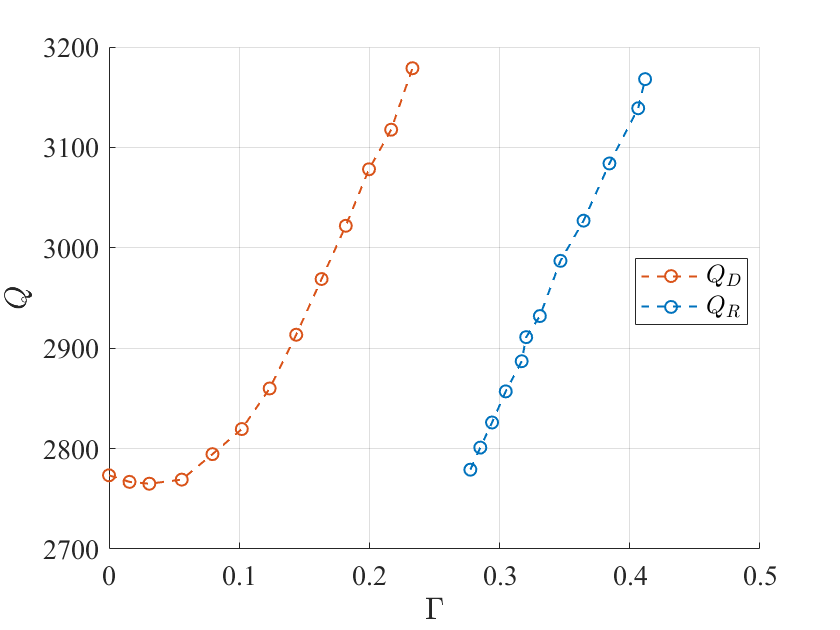}
    \caption{Loaded quality factor of the DDFB cavity (orange dots) and of the DFB cavity (blue dots) for $N=60$ unit cells, against waveguide end-reflection coefficient $\Gamma$ at each of the four terminations. The DDFB cavity does not require waveguide end-reflection coefficient at its terminations. Even with vanishing end-reflection coefficient, the DDFB cavity has a $Q$ comparable with that of the DFB cavity with higher end-reflection coefficient. Note that the DDFB cavity is more resilient to perturbations of its terminations than the DFB cavity.} 
    \label{fig:QvsRefl}
\end{figure}

\subsection{Effect of End Reflections on Cavity Resonances}
The quality factors of the DDFB and DFB cavities can be controlled to some degree by resorting to the concept of waveguide impedance. Figure~\ref{fig:QvsRefl} depicts the change in the loaded quality factors of the DDFB (orange dots) and DFB cavities (blue dots) with $N=60$ as a function of the  "termination parameter" $\Gamma$ of the individual waveguides, at their respective DBE and RBE resonances. Only the case with a relatively small $N=60$ is shown here to limit the computational burden. 

For simplicity, and without losing generality, in this paper  we describe the impedance mismatch of the individual waveguides using the field-reflection termination parameter

\begin{equation}
    \Gamma = \frac{\eta_t - \eta_w}{\eta_t + \eta_w},
    \label{eq:Gamma}
\end{equation}

where $\eta_w$ and $\eta_t$ are the wave impedances of the cavity waveguides and outside waveguides (used as ``terminations''), respectively. We approximate $\eta_w \approx \sqrt{\mu_0/\epsilon_w}$ where $\epsilon_w$ is the {\em effective} permittivity of the even mode of the two straight coupled waveguides inside the DDFB cavity, without the corrugations. Similarly, $\eta_t \approx \sqrt{\mu_0/\epsilon_t}$, where $\epsilon_t$ is the {\em effective} permittivity of the even mode of the two straight coupled waveguides outside the DDFB cavity, used as termination. We use the $\varepsilon_w$ of the even mode of the coupled waveguides, although using the odd mode effective permittivity would yield similar results. 

 We emphasize that $\Gamma$ is a single-waveguide {\em termination parameter} that approximately characterizes the additional reflection introduced by the impedance mismatch between the straight waveguide sections at the cavity ends. It should not be confused with the actual complex reflection coefficients experienced by the Bloch modes of the periodic double-grating structure. Indeed, near the DBE, the electromagnetic field involves four Bloch modes, including propagating and evanescent components, whose coupling at the cavity boundaries cannot be described by a single scalar reflection coefficient. A rigorous description would require a multimode scattering formulation, involving a reflection matrix and the Bloch impedances of the periodic structure (see \cite{othman_theory_2016} for details on Bloch impedance behavior near EPDs). In particular, $\Gamma=0$ indicates that the straight waveguide sections have no additional impedance discontinuity, but does not imply the absence of reflection at the boundaries of the periodic cavity. Such intrinsic reflection arises from the mismatch between the Bloch modes of the periodic structure and the modes of the external uniform waveguides, and remains present even when $\Gamma=0$. Importantly, the full-wave FEM simulations account for these effects through the actual geometry and electromagnetic boundary conditions, without imposing the approximate parameter $\Gamma$ as a reflection boundary condition. The same definition of $\Gamma$ is adopted for the reference DFB cavity with $s=0$. In summary, the termination parameter $\Gamma$ provides a convenient measure of the additional mirror effect introduced by the impedance mismatch between the straight waveguide sections at the cavity ends. 
In particular, $\Gamma=0$ corresponds to a mirrorless cavity, where the periodic double-grating structure is connected directly to uniform waveguides without additional mirrors. Here, “mirrorless” indicates that optical feedback is provided by the periodic structure itself, without requiring dedicated end mirrors; it does not imply the absence of intrinsic reflections associated with Bloch-mode impedance mismatch at the boundaries of the periodic structure.

In our simulations, the change in $\Gamma$ is implemented by modifying the refractive index of the external waveguides' core, in the sections with length $L=1.2$ $\upmu$m on each side of the cavity. For the DDFB cavity, the distance between the last corrugation and the refractive index discontinuity (shown by two outermost vertical black dashed lines in Figure~\ref{fig:DDFBStructure}) is approximately $67$ nm. For the DFB cavity, the distance is $127$ nm. In practice, a distributed Bragg reflector (DBR) would be used to tailor the reflectivities \cite{ghafouri_distributed_2003_ch1} at the edges of the cavity, but here we prefer to use a simpler implementation to focus on the physics results associated with the cavity termination effect rather than on the design of a mirror. Since the DDFB cavity displays significantly larger quality factors than a DFB cavity of the same length, as seen in Figure~\ref{fig:QvsN}, to achieve the same quality factor, the DDFB cavity requires much lower {\em additional end-reflection coefficient} $\Gamma$ than a DFB cavity of the same length. As shown in Fig.~\ref{fig:QvsRefl}, no end-reflection coefficient is needed for the DDFB cavity to have a large $Q$ factor. Indeed, the quality factor of the DDFB cavity without  end-reflection coefficient, i.e., with $\Gamma = 0$, is comparable with that of the DFB cavity of the same length with a higher end-reflection coefficient, $\Gamma = 0.3$. Additionally, the DDFB cavity is also shown to be more resilient to variations in the end-reflection coefficient $\Gamma$ of its terminations than the DFB cavity, especially around the mirrorless $\Gamma=0$ case, where $Q_D$ varies with $\Gamma$ and forms a small parabola. One of the major conclusions is that there is a large reflection at the two DDFB cavity terminations even when the cavity is made mirrorless, $\Gamma=0$, i.e., when the two chirped waveguides are connected to two straight waveguides with the same thickness $t$ and material and without corrugations. 

\begin{figure}
    \centering
    \begin{subfigure}{0.48\textwidth}
        \includegraphics[width=\linewidth]{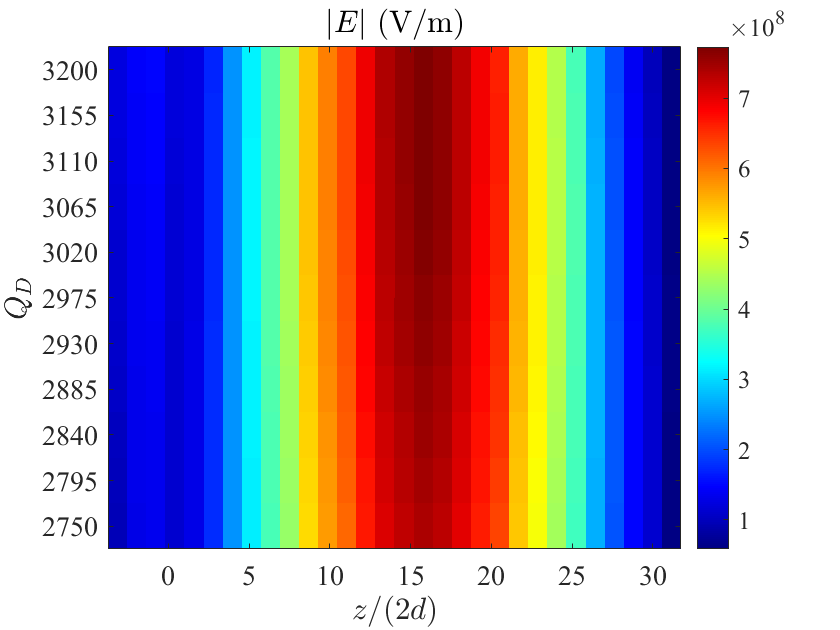}
        \caption{DDFB cavity}
    \end{subfigure}
    
    \begin{subfigure}{0.48\textwidth}
        \includegraphics[width=\linewidth]{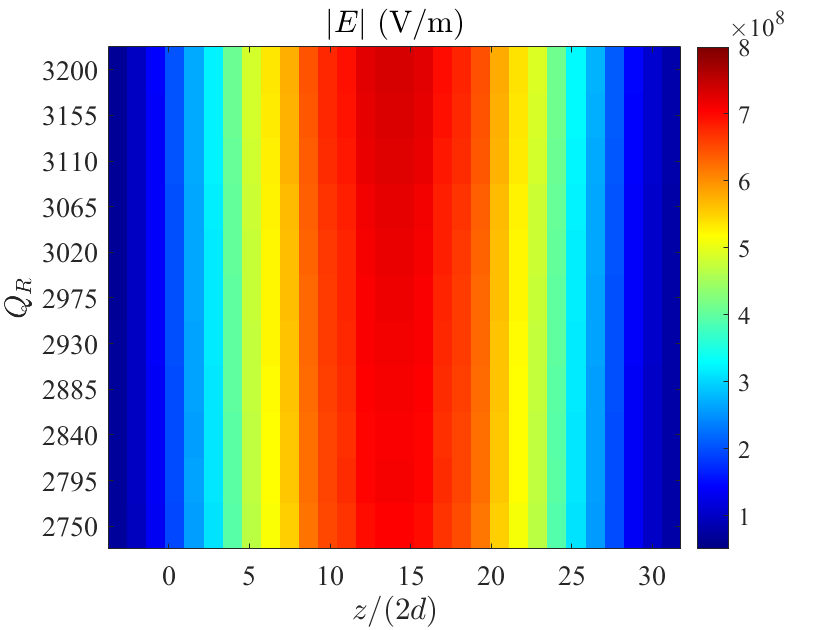}
        \caption{DFB cavity}
    \end{subfigure}
    \caption{ Spatial distributions of the electric-field magnitude in the passive (a) DDFB and (b) DFB cavities with $N=60$, evaluated at their respective DBE and RBE resonances. The different loaded quality factors are obtained by varying the additional end-reflection parameter $\Gamma$ as shown in Fig.~\ref{fig:QvsRefl}. Each pixel represents the electric-field magnitude averaged over two consecutive unit cells along $z$ and over a vertical region of thickness $t+50$ nm encompassing the upper waveguide. The DDFB cavity exhibits stronger field enhancement than the DFB cavity under the considered conditions.}
    \label{fig:Qfields}
\end{figure}

 Although Fig.~\ref{fig:QvsRefl} illustrates the dependence of the quality factor on the additional end-reflection coefficient for $N=60$, the length dependence of the mirrorless cavity ($\Gamma=0$) has already been investigated in Fig. 4. The numerical results in Fig. 4 are consistent with the asymptotic $Q_D\propto N^5$ scaling over the investigated range of cavity lengths, without requiring additional end mirrors.
Figure~\ref{fig:Qfields} shows the spatial electric-field profiles at the respective resonances of the (a) DDFB and (b) DFB cavities, each consisting of $N=60$ unit cells, for different loaded quality factors obtained by varying the additional end-reflection parameter $\Gamma$, as shown in Fig.~\ref{fig:QvsRefl}. The plotted fields in both figures correspond to those in the top waveguide, including the evanescent fields extending up to $25$ nm above and below the core. The value of each cell in Figure~\ref{fig:Qfields} is determined as the average of the field magnitude over a box of total vertical thickness $t+50$ nm, covering two consecutive unit cells along the $z$ direction (the full extent of the corrugation is not included within the box). The strong field amplitudes away from the cavity's edge at the DBE resonance, especially compared to the field amplitudes at the RBE resonance for the same quality factor, demonstrate a uniquely "structured" resonance field that leads to an enhancement of the local density of states \cite{othman_giant_2016}. We observe a tight field confinement inside the DDFB waveguide at the DBE resonance that is consistent with the results reported in Ref.~\cite{mealy_degenerate_2023}. The large relative field magnitude inside the waveguide further confirms the DBE-associated frozen-mode behavior, especially compared with the fields in the DFB cavity of the same length and quality factor. The value of field enhancement in a cavity is markedly reliant on the topology, technology, and length of the optical waveguides (for instance, a different waveguide supporting the DBE reported in Ref.~\cite{nada_giant_2018} provides a different field enhancement than the illustrative case provided here). In essence, the maximum field enhancement in a cavity with a DBE is generally larger than that of a comparable cavity with an RBE having the same $Q$ factor, length, and analogous topologies. This is also seen in Ref.~\cite{veysi_degenerate_2018}, and such an observation can be generalized to other implementations of cavities supporting a DBE and an RBE. Here we have shown the field enhancement for $N=60$. However, the maximum field intensity in a DDFB cavity is expected to scale as $N^4$ near a DBE as demonstrated in  previous pioneering work on DBE structures \cite{figotin_gigantic_2005}, whereas the field enhancement in a DFB cavity should scale as $N^2$. Also the {\em local density of states} has been demonstrated to follow the $N^4$ asymptotic scaling law in  \cite{othman_giant_2016} in a previous case of a DBE cavity.

These results are important because they demonstrate the superior cavity effect of the DBE-associated frozen mode compared to that of the RBE-associated standing wave. 

In summary, the DDFB cavity displays a larger quality factor than a DFB cavity of the same length, displaying a larger scaling power with waveguide length ($Q_D\sim N^5$ as opposed to $Q_R\sim N^3$). When the additional end-reflection coefficient of their terminations is tailored, the DDFB cavity necessitates lower waveguide reflectivities to have a $Q$ of the same value as that of the DFB cavity. These results have been shown here for $N=60$, but the difference in end-reflection coefficient to provide the same $Q$ is even starker for larger $N$, owing to the $N^5$ and $N^3$ asymptotic scaling laws at the DBE and RBE resonances. 
Analogously, the difference between the maximum field enhancements in the two cavities has been shown only for $N=60$ to simplify the numerical burden, but their difference is even stronger for larger $N$, owing to the $N^4$ and $N^2$ asymptotic trends for field enhancement at DBE and RBE resonances, respectively, as shown in Ref. \cite{figotin_gigantic_2005}. Additionally, the quality factor of the DDFB cavity is more resilient to perturbations of the end-reflection coefficient of its terminations than the quality factor of the DFB cavity.

We have also shown that a DDFB cavity exhibits stronger fields away from the cavity's edges compared to a DFB cavity of the same length and quality factor. This happens with approximately half the end-reflection coefficient at the DDFB cavity edges than that of the DFB cavity when $N=60$, with a starker difference expected for larger $N$.

\section{DDFB-induced lasing threshold}
\label{ch:Lasing}

\subsection{Gain Model and Lasing Threshold Determination}
The goal of this section is to show that the lasing threshold of a DDFB cavity is lower than that of a DFB cavity and determine the scaling law of the DDFB threshold versus grating length. To do so, we consider the gain to be uniformly distributed in the waveguide cladding (i.e., in the bulk) \cite{yariv_photonics_2007, parareda_lasing_2023}. There are various ways to add gain, including rare-earth material doping, current injection, and the use of quantum wells \cite{siegman_lasers_1986}, which increase the field amplitude of the waves traveling in the cavity. Current injection is commonly used to add gain in DFB lasers \cite{ghafouri_distributed_2003_ch2}. 

Here, we describe the uniform distribution of gain in the cladding as the imaginary part $-n_g''$ of the complex refractive index $n=n_{b}-j n_g''$, where $n_b$ is the bulk refractive index, and gain corresponds to $n_g'' <0$. Although the FEM solver used here (CST Studio Suite) does not directly support a negative imaginary part of the refractive index, the solver allows for a complex relative permittivity $\varepsilon$. Using the procedure described in \cite{gent_CSTgain_2011}, we introduce gain to the simulation. 
 Representing optical gain through the complex relative permittivity is electromagnetically equivalent to specifying a complex refractive index, since $\varepsilon=n^2$ for a nonmagnetic isotropic medium. Therefore, the use of complex permittivity in the FEM solver does not introduce an additional approximation in the representation of gain at the operating frequency.
The complex permittivity of the cladding, $\varepsilon$ is written as $\varepsilon = \varepsilon' - j\varepsilon''$, so its imaginary part is $- \varepsilon'' = \text{Im}(n^2) = -2n_b n_g''$.  The approximation arises only when the quadratic contribution of the imaginary refractive-index component is neglected in the conversion formula $\varepsilon^\prime
=n_b^2-(n_g'')^2\approx n_b^2$.
 
The conversion between the bulk gain coefficient $\alpha_g$ (in units of $\text{dB}/\text{cm}$) and the gain in terms of the imaginary part of a refractive index at an angular frequency $\omega$ is approximated by $n_g''(\omega) \approx -\alpha_g (100/8.686)(c_0/\omega)$ where $c_0$ is the speed of light in vacuum. 

The gain region includes the cladding between the two gratings, which increases the interaction between the waves and the gain medium \cite{ghafouri_distributed_2003_ch2}. 
 The uniform gain distribution considered here is an idealized model intended to isolate the effect of the DBE on the lasing threshold, rather than to represent a specific electrically pumped semiconductor laser. In a practical implementation, gain may be confined to selected regions, such as quantum wells or quantum dots, and the lasing threshold will depend on the spatial overlap between the resonant electromagnetic field and the active material. A smaller overlap generally requires a larger material gain to compensate for the cavity losses. Consequently, the threshold values reported here are specific to the assumed gain distribution. Nevertheless, the characteristic inverse-fifth-power scaling with cavity length is expected to persist when the effective gain overlap remains approximately independent of cavity length and the DBE resonance is preserved.

We define the lasing threshold $\alpha_{th}$ as the minimum amount of bulk gain necessary for an active cavity to start oscillating due to the generation of an instability \cite{siegman_lasers_1986,ghafouri_distributed_2003_ch1, veysi_degenerate_2018}. 

 The lasing threshold is generally inversely related to the passive-cavity quality factor, $\alpha_{\mathrm{th}}\propto 1/Q$ [67], provided that the effective spatial overlap between the gain medium and the resonant electromagnetic field remains approximately independent of cavity length and that additional length-dependent losses do not dominate. Since the DBE quality factor scales asymptotically as $Q_D\propto N^5$ \cite{figotin_slow_2011, wood_degenerate_2015, othman_giant_2016, nada_theory_2017, othman_experimental_2017}, the lasing threshold is expected to follow $\alpha_{D,\mathrm{th}}\propto N^{-5}$. This threshold scaling has previously been observed numerically in idealized DBE-based cavity models \cite{othman_giant_2016, veysi_degenerate_2018}. In the following, we investigate whether the same scaling law is obtained for the proposed double-grating photonic structure using full-wave electromagnetic simulations.

To determine the lasing threshold of a finite-length DDFB cavity, we gradually increase the bulk gain while monitoring the cavity field for unstable behavior. This is done by tracking the poles of the $S_{21}$ transmission parameter at the DBE resonance, as previously shown in  Appendix~A of \cite{nada_exceptional_2020}, Appendix~B of \cite{kessem_low_2023} or in \cite{furman_impact_2025} for other kinds of exceptional degeneracies. To ensure that lasing at a resonance $f_{r,D}$ near the DBE frequency $f_D = 193.27$ THz is preferred over lasing near the closest RBE at $189.38$ THz in the same cavity (the blue curve in Fig.~\ref{fig:DispDiagr}), we computed the lasing thresholds at the nearest resonances to the two exceptional points for $N=60$ and $N=120$ of the same structure. For $N=60$, the lasing threshold at the DBE resonance (the closest one to the DBE frequency) is $\alpha_{D,th} = 505$ dB/cm, while that near the RBE resonance (closest to the RBE frequency) is $\alpha_{th}=2390$ dB/cm, resulting in a $4.7$ times larger value than the DBE one. For $N=120$, we have that $\alpha_{D,th} = 40$ dB/cm and the threshold at the RBE resonance is $\alpha_{th} = 295$ dB/cm (that is $7.4$ times larger than the DBE one). The lasing thresholds at resonances associated with the DBE are significantly lower than those associated with the nearest RBE. Therefore, in the double-grating cavity of Fig.~\ref{fig:DDFBStructure} that supports both the DBE and the RBE, lasing near the DBE frequency is preferred over lasing at the nearest RBE resonance. 

\begin{figure}
    \centering
    \includegraphics[width=0.42\textwidth]{pictures/Gratings_Fig3_7_v2.png} 
    \vspace{2mm} 
    \includegraphics[width=0.5\textwidth]{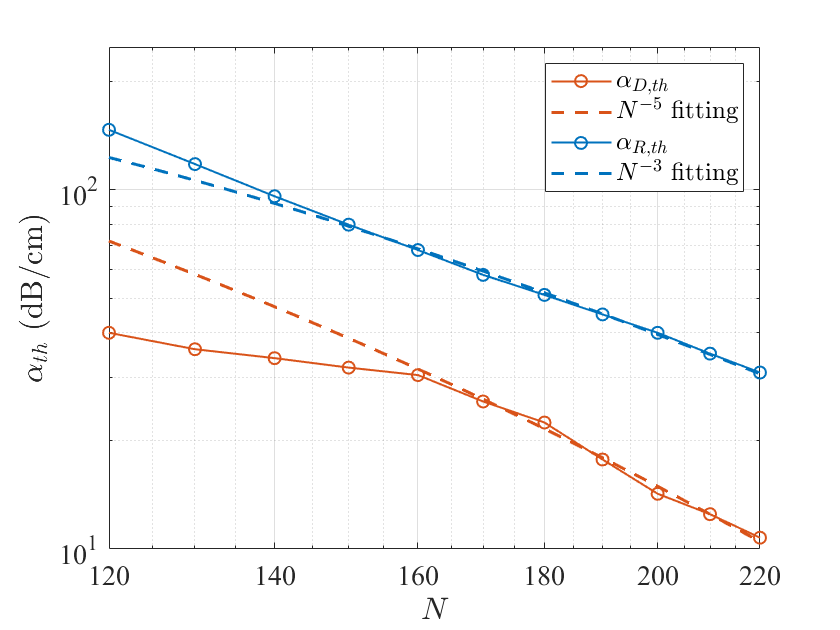}
    \caption{ Lasing thresholds of the DDFB (orange dots) and DFB (blue dots) cavities versus the number of unit cells $(N)$, evaluated at their respective DBE and RBE resonances. The dashed curves represent fits to the numerical results, illustrating the asymptotic threshold scaling laws $\alpha_{D,\mathrm{th}}\propto N^{-5}$ and $\alpha_{R,\mathrm{th}}\propto N^{-3}$, respectively.}
    \label{fig:ThvsN}
\end{figure}

\subsection{Lasing Threshold Scaling and Termination Effects}
We have just compared the DBE and RBE lasing threshold in the same cavity, assuming first $N=60$ and then $N=120$. Now we compare the thresholds in two different cavities made of two different double gratings: the DDFB one (with $s\ne 0$) and the DFB one (with $s=0$). The DDFB cavity  supports a DBE at $f_D$, and has threshold $\alpha_{D,th}$, whereas the DFB cavity is designed to support an RBE at $f_R \approx f_D$ (shown in Fig.~\ref{fig:DispDiagr} with a dashed-yellow line), and has threshold $\alpha_{R,th}$. The two thresholds are shown respectively as orange and blue dots in Figure~\ref{fig:ThvsN} for different grating lengths $N$. The DDFB cavity lasing threshold is always smaller than the DFB one, for any $N$ in the simulated range. The tapering in the cavities is the same as in the results depicted in Figure~\ref{fig:QvsN}. 
\begin{table}[t]
\centering
\caption{Comparison of the band-edge degeneracy order
and asymptotic scaling laws for DDFB and DFB cavities.}
\label{tab:DDFB_DFB}
\renewcommand{\arraystretch}{1.2}
\begin{tabular}{lcc}
\hline
Property & DDFB (DBE) & DFB (RBE) \\
\hline
Degeneracy order & 4 & 2 \\
Quality factor
    & $Q_D \propto N^5$
    & $Q_R \propto N^3$ \\
Lasing threshold
    & $\alpha_{D,\mathrm{th}} \propto N^{-5}$
    & $\alpha_{R,\mathrm{th}} \propto N^{-3}$ \\
\hline
\end{tabular}
\end{table}
To determine the asymptotic scaling laws of the threshold for the two DDFB and DFB cavities versus grating lengths, we fit the threshold found numerically with two fitting functions, in orange and blue dashed lines, respectively. For the DDFB cavity, the trend is established by fitting $\alpha_{D,th}(N)$ with the function 

\begin{equation}
\alpha_{D,th}^{fit}(N) = 1/(a N^5+b N^2),
    \label{eq:LasThreshDDFB}
\end{equation}

and we calculate the fit over $N \in (160, 200)$. The fitting coefficients are calculated as $a=1.29\times 10^{- 13}$ and $b=6.78 \times 10^{- 7}$. For $N>173$, the $aN^{5}$ term dominates the denominator, yielding the asymptotic threshold scaling $\alpha_{D,\mathrm{th}}\propto N^{-5}$. We also verified that the difference $N^5 \times|\alpha_{D,th}^{fit}-\alpha_{D,th}|$ approaches a constant value for large $N$.  The numerical results are consistent with the asymptotic scaling
\begin{equation}
\alpha_{D,th} \propto 1/ (aN^{5})
\end{equation}
as indicated by the fitted trend and its agreement with the calculated thresholds over the investigated range of cavity lengths.

The trend for the lasing thresholds versus DFB cavity length is established in a similar manner, leading to the fitting function 

\begin{equation}
\alpha_{R,th}^{fit}(N) = 1/(cN^3+d), 
    \label{eq:LasThreshDDFB}
\end{equation}

with $c=2.85\times 10^{-9}$ and $d=3 \times 10^{-3}$. Since the $c N^{3}$ term dominates the denominator for $N > 101$, this result shows the asymptotic scaling threshold $\alpha_{R,th} \propto 1/( cN^{3})$ for large $N$. 

 The quality-factor scaling laws obtained from our full-wave simulations are consistent with established theoretical predictions for finite-length cavities operating near regular and degenerate band edges. Similarly, the observed lasing-threshold scaling agrees with previous numerical studies based on idealized coupled-mode models. This agreement across different modeling approaches supports the physical origin of the observed scaling laws and provides additional confidence in the numerical results.

The main implications of these trends are twofold. Primarily, the mirrorless DDFB cavity displays significantly lower lasing thresholds than a mirrorless DFB cavity of the same length. For instance, the DDFB and DFB cavities of $N=60$ unit cells without mirrors exhibit a quality factor of $Q_D = 3451$ and $Q_R = 1378$, and a lasing threshold of $\alpha_{D,th} = 505$ and $\alpha_{R,th} = 1099$,  respectively. The mirrorless DDFB cavity requires less than half the amount of gain (in $\text{dB/cm}$) to start oscillation of the preferred mode than a mirrorless DFB cavity of the same length for $N=60$.  By increasing $N$ to only $N=120$, $\alpha_{D,th} = 40$ dB/cm while $\alpha_{R,th} = 147$ dB/cm, and this difference increases further for larger $N$. Because of the different asymptotic lasing threshold laws for the DDFB cavity, $\alpha_{D,th} \propto N^{-5}$, and the DFB cavity, $\alpha_{R,th} \propto N^{-3}$, the DDFB lasing principle is characterized by a different mechanism than that of a DFB laser. This suggests possibilities for device miniaturization and reduced power consumption. Indeed, when the trends are established for large-enough $N$, the ratio of the thresholds between the two cavities decreases asymptotically as 

\begin{equation}
\frac{\alpha_{D,th}}{\alpha_{R,th}}
 \sim 
 \frac{c}{a}\frac{1}{N^{2}}. 
    \label{eq:RatioThresholds}
\end{equation}

\begin{figure}
    \centering
    \includegraphics[width=0.48\textwidth]{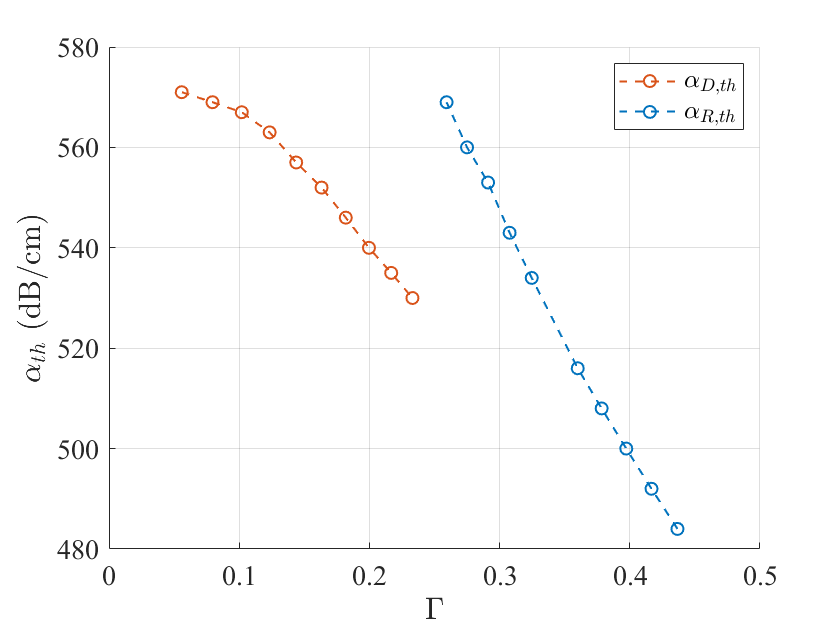}
    \caption{Comparison between the lasing thresholds of the DDFB (orange dots) and the DFB (blue dots) cavities with the same length ($N=60$) as a function of the additional end-reflection coefficient $\Gamma$ at their terminations. The DDFB cavity necessitates much lower reflectivities than the DFB cavity of the same length to display the same lasing threshold. Additionally, the threshold of the DDFB cavity is more resilient to variations in its terminations than the DFB cavity.}
    \label{fig:ThvsRefl}
\end{figure}

Figure~\ref{fig:ThvsRefl} depicts the lasing thresholds of the DDFB (orange dots) and the DFB (blue dots) cavities with the same length $N=60$ as a function of $\Gamma$ as defined in Eq.~(\ref{eq:Gamma}). Primarily, it shows that the lasing threshold of the DDFB cavity is more resilient to changes in its terminations than that of the DFB cavity of the same length and quality factor. This trend is enhanced when the waveguides are long enough so that the asymptotic regimes for the scaling of the quality factor and lasing threshold at DBE resonances are established.

\subsection{Practical Considerations and Limitations}

The anomalous scaling laws $Q_D\propto N^5$ and $\alpha_{D,\mathrm{th}}\propto N^{-5}$ are obtained for an ideal periodic structure and may be limited by fabrication imperfections \cite{nada_giant_2018}.  In particular, deviations in the refractive indexes, grating dimensions, separation, and relative longitudinal shift may perturb the quartic dispersion associated with the DBE. Systematic deviations can shift or split the degeneracy, whereas random cell-to-cell variations may introduce additional scattering and modify the cavity resonances. As the number of unit cells increases, these effects can become increasingly important and eventually prevent the ideal scaling laws from being maintained. Nevertheless, moderate perturbations do not necessarily eliminate the enhanced degenerate-resonance properties associated with operation near a DBE. The quantitative tolerance limits depend on the specific waveguide geometry and fabrication process and are not determined in this work.

 Experimental studies of passive silicon photonic waveguides have demonstrated resonance properties associated with the DBE \cite{wood_degenerate_2015,reano_exp_2016}, supporting the feasibility of fabricating periodic structures designed to operate near a fourth-order EPD. However, lasing in the DDFB configuration proposed here has not yet been experimentally demonstrated. A practical implementation would require the integration of an active material with sufficient optical gain and spatial overlap with the DBE-associated resonant field. Possible approaches include incorporating semiconductor gain regions, such as quantum wells or quantum dots, or integrating a suitable optically pumped gain medium. Electrical pumping would additionally require an appropriate current-injection geometry and electrical contacts that preserve the desired optical coupling and periodic dispersion. These considerations, together with fabrication tolerances, optical losses, and thermal effects, may affect the achievable lasing threshold. The present work therefore establishes the underlying lasing mechanism and its idealized threshold properties, while the design and experimental realization of a practical electrically or optically pumped DDFB laser remain subjects for future investigation.
Overall, we have shown that mirrorless DDFB cavities display much larger quality factors and much lower lasing thresholds than comparable mirrorless DFB cavities of the same length. The quality factor and lasing thresholds of the DDFB cavity scale with waveguide length as the fifth power and inverse fifth power, respectively. In the DFB cavity, they scale as $N^3$ and $N^{-3}$ instead. A related discussion in \cite{othman_giant_2016} regarding the local density of states distribution within the cavity for the DBE and RBE cases supports these claims. Additionally, we have shown that the DDFB cavity displays larger field amplitudes than the DFB cavity even when their loaded $Q$ factors are the same. The lasing thresholds of both cavities can be tuned by adjusting the reflectivities at the cavities' ends. Importantly, the DDFB cavity has also been shown to be significantly more resilient to perturbations of the terminations of the lasing cavity, demonstrating its robustness to external effects.   
The fundamental differences between the DDFB and DFB cavities, in terms of band-edge degeneracy order and the asymptotic scaling laws of their quality factors and lasing thresholds, are summarized in Table~\ref{tab:DDFB_DFB}.

Though not shown here, the resonance frequencies at which the thresholds are calculated are very close to the DBE resonances $f_{r,D}$ that, by virtue of (\ref{eq:DBEresFreqAsymp}), are asymptotically close to the DBE frequency $f_{D}$ for large $N$. This means that by increasing the length of the grating, the oscillation frequency should be stable and close to $f_D$. Furthermore, the lasing frequency should be resilient to changes in load terminations. Indeed, this phenomenon has been already demonstrated in different but highly related studies at radio frequency accounting for nonlinear saturation effects, theoretically and experimentally, in Refs.  \cite{oshmarin2019IET-newOsc, Abdleshafy2020TAP-DistrDBEoscc,Oshmarin2021arXiv-experDBEoscillator} by varying the cavity loading.

\section{Conclusion}
\label{ch:Conclusion}
We have investigated degenerate distributed feedback (DDFB) lasing in a realistic double-grating photonic cavity operating near a fourth-order degenerate band edge (DBE) hat is a special exceptional point of
order four in a lossless and gainless multimode waveguide. We have also provided an analytical demonstration that the  quartic flat Bloch dispersion of a four-mode system implies a single fourth-order Jordan block, corresponding to the coalescence of all four Bloch eigenvectors. This establishes the mathematical connection between the characteristic DBE dispersion and the fourth-order EPD without requiring explicit eigenvector calculations. Using full-wave electromagnetic simulations, we have demonstrated that the DDFB cavity exhibits an asymptotic quality-factor scaling $Q_D\propto N^5$ and a lasing-threshold scaling $\alpha_{D,\mathrm{th}}\propto N^{-5}$, compared with $Q_R\propto N^3$ and $\alpha_{R,\mathrm{th}}\propto N^{-3}$ for a comparable DFB cavity operating near a regular band edge (RBE). These results demonstrate the advantages of the four-mode degeneracy for enhancing optical feedback and reducing the lasing threshold.

The DDFB cavity also exhibits stronger intracavity field enhancement than the reference DFB cavity, even for comparable loaded quality factors. Furthermore, it achieves lower lasing thresholds without additional end mirrors and shows reduced sensitivity to variations in the termination reflectivity. These properties highlight the potential advantages of the DBE-based lasing mechanism over conventional RBE-based distributed feedback.

As the cavity length increases, the DBE resonance frequency approaches the DBE frequency, suggesting the potential for stable, single-frequency operation. The proposed DDFB concept may therefore offer opportunities for compact, low-threshold lasers for optical communications, sensing, and spectroscopy. Its implementation in different photonic material platforms represents a promising direction for future investigation.

\appendix
Consider the characteristic equation of the unit-cell $4\times 4$ transfer matrix
\begin{equation}
D(\zeta,\omega)=\det[\mathbf {\mathrm T}_{ \mathrm U}(\omega)-\zeta\mathbf I]=0,
\qquad \zeta=e^{-jkd}.
\end{equation}

At the DBE, $k=k_D \equiv \pi/d$, and therefore $\zeta=\zeta_D \equiv-1$.  The quartic dispersion near the DBE is
\[
\omega_D-\omega\simeq h(k-k_D)^4,\qquad h\ne0.
\]
Since $\zeta=e^{-jkd}$, its expansion near $k_D$ gives $\zeta+1\simeq jd(k-k_D)$.
Consequently, the quartic dispersion can be written locally as
\begin{equation}
\omega-\omega_D=C(\zeta+1)^4+\cdots,
\qquad C\ne0.
\label{eq:quarticDisp}
\end{equation}
In a waveguide system made of four modes, two in each direction, the transfer matrix is $4\times4$, and its characteristic equation is a fourth-degree polynomial in $\zeta$. At the DBE, all four Bloch eigenvalues coincide at $\zeta_D=-1$, giving an eigenvalue of algebraic multiplicity four. Assuming an analytic transfer matrix and the nondegenerate quartic dispersion in Eq.~\eqref{eq:quarticDisp}, the local characteristic equation has the leading-order expansion
\begin{equation}
D(\zeta,\omega)
=A(\zeta+1)^4+B(\omega-\omega_D)+\cdots,
\end{equation}
where $A\ne0$ and $B\ne0$. Indeed, if $B=0$, the characteristic equation would not produce the four distinct fourth-root branches associated with the nondegenerate quartic dispersion. The condition $B=\left.\frac{\partial D}{\partial\omega}\right|_{\mathrm{DBE}}\ne0$  
together with the fourfold algebraic multiplicity of $\zeta_D=-1$, implies that the transfer matrix has a single fourth-order Jordan block, as established by Theorem 2.1 of \cite{welters_explicit_2011}. For completeness, we provide here a direct proof that the fourfold eigenvalue $\zeta_D=-1$ must have only one independent eigenvector.
Suppose, by contradiction, that it has two or more independent eigenvectors. Then
\begin{equation}
\dim\ker[\mathbf T_{ \mathrm U}(\omega_D)+\mathbf I]\ge2.
\end{equation}
Since $\mathbf T_U$ is a $4\times4$ matrix, the rank-nullity theorem implies 
\begin{equation}
\operatorname{rank}[\mathbf T_U(\omega_D)+\mathbf I]\le2.
\end{equation}

For a $4\times4$ matrix $\mathbf M$, the adjugate matrix $\operatorname{adj}(\mathbf M)$ is the transpose of the cofactor matrix, whose elements are the signed determinants of the $3\times3$ submatrices obtained by deleting one row and one column of $\mathbf M$. Since $\operatorname{rank}(\mathbf M)\le2$, at most two columns of $\mathbf M$ are linearly independent. Consequently, any three columns are linearly dependent, implying that the determinant of every $3\times3$ submatrix vanishes. All cofactors are therefore zero, and hence $\operatorname{adj}(\mathbf M)=0$. Applying this result to $\mathbf M=\mathbf T_{ \mathrm U}(\omega_D)\mathbf+I$, we obtain

\begin{equation}
\operatorname{adj}[\mathbf T_{ \mathrm U}(\omega_D)+\mathbf I]=0.
\end{equation}

Using Jacobi's formula for the derivative of a determinant \cite{bellman1997introductionMatrix},
\begin{equation}
\frac{\partial D}{\partial\omega} = 
\operatorname{tr}\left[
\operatorname{adj}(\mathbf T_{ \mathrm U}-\zeta \mathbf I)
\frac{\partial \mathbf T_{ \mathrm U}}{\partial\omega}
\right],
\end{equation}
we obtain, at the DBE, $\left.\frac{\partial D}{\partial\omega}\right|_{\mathrm{DBE}}=0$. This contradicts the condition $B\ne0$ obtained from the local characteristic expansion.
Therefore,
\begin{equation}
\dim\ker[\mathbf T_{ \mathrm U}(\omega_D)+\mathbf I]=1.
\end{equation}
Since the eigenvalue $\zeta_D=-1$ has algebraic multiplicity four but geometric multiplicity one, the transfer matrix has a single fourth-order Jordan block. Thus, all four eigenvectors coalesce into one eigendirection, establishing the connection between the quartic DBE dispersion and a fourth-order EPD.

\section*{Acknowledgment}
The authors are thankful to DS SIMULIA for providing CST Studio Suite.
\ifCLASSOPTIONcaptionsoff
  \newpage
\fi

\bibliographystyle{IEEEtran}
\bibliography{MAIN_DDFB_lasing}

\end{document}